\documentstyle[11pt]{article}
\begin{document}
\begin{flushright}
UCTP-105/97
\end{flushright}
\begin{center}
{\Huge Solution of gauge theories induced by fundamental representation
scalars}
\\
\vskip0.5in
P. Suranyi\\ University of Cincinnati,
Cincinnati, Ohio 45221
\vskip0.5in
{\bf Abstract}
\end{center}
\vskip0.1in
\noindent Gauge theories induced by scalars in the fundamental representation
of the $U(N_c)_{\rm gauge}\times U(N_f)_{\rm global}$ group are
investigated in the large $N_c$ and $N_f$ limit.  A master field is defined
from bilinears of the scalar field following an Eguchi-Kawai type reduction of
spacetime. The density function for the master field satisfies an integral
equation that can be solved exactly in two dimensions ($D=2$) and in a
convergent series of approximations at $D>2$.  While at $D=2$ the system is in
the same phase at all $\epsilon=N_c/N_f$, it undergoes a phase transition at a
critical value,  $\epsilon_c(D)$, for $D>2$.

 \vskip0.3in

\section{Introduction}
Many quantum field theories, with the notable exception of nonabelian gauge
theories, can be solved exactly when the number of fields, $N$, tends to
infinity.
 The large $N$ expansion is not overly useful for solving nonabelian gauge
theories   because the leading order of the  expansion  is equivalent to the
sum
of all planar diagrams. Unfortunately, it is not much
simpler to sum planar diagrams than to solve the exact theory.

To obtain exact results further simplifications, like those proposed by
Kazakov and Migdal~\cite{kazakov}, are required.  Using the lattice action of a
gauged adjoint representation scalar field, in which the pure gauge term was
omitted, these authors derived an equation for the master field, composed of
the
eigenvalues of the scalar field matrix.  Once the master field is known
other physical parameters of the system can also be calculated. Following the
philosophy that nonabelian gauge theories are unique one hopes  that correct
self-interactions of the gauge field are generated by scalar dynamics.

  There are two reasons why  gauge theories induced by adjoint
representation scalars ~\cite{kazakov}~\cite{gross} may not be completely
equivalent to standard nonabelian theories. The first reason is well-known:
Because
 the action is quadratic in the compact gauge field, $U$, there is a hidden
local $Z_N$ symmetry, which implies that the vacuum
expectation value of Wilson loops is zero.~\cite{semenoff}

The second reason which makes one believe that induced theories
 and standard theories are different, goes as follows. Suppose
that along with Araf'eva~\cite{arefeva} we gauge only one of the indices of
the adjoint representation scalar field. Then the theory has a $U(N_c)_{\rm
gauge}\times U(N_c)_{\rm global}$ symmetry and the unwanted $Z_N$
symmetry is not present. We can view such a theory as one having  global and
gauge symmetries with equal numbers of flavors and colors.
The gauge theory induced by such a scalar field is not asymptotically free in
four dimensions and as such is not likely to be confining either. In fact, the
induced theory will only be asymptotically free if the number of complex
scalars, $N_f$, satisfies the inequality
$N_f>11N_c$~\cite{hasenfratz},\cite{pap}.
  This is the
primary reason why we previously considered a scalar field theory with
$U(N_c)_{\rm gauge}\times U(N_f)_{\rm global}$ symmetry.~\cite{pap}

The prevalent argument against theories with fundamental representation
scalars,
rather then adjoint representation ones has been the lack of existence of a
master field, precluding chances for finding exact solutions.
While in gauge theories induced by adjoint
representation scalars the master field is composed of the eigenvalues of
the scalar field matrix, for
fundamental representation scalars no such quantity seems to exist.

The purpose of this paper is to show that the general bias against gauge
theories induced by fundamental representation scalars is unfounded. It is
possible to introduce an adjoint gauge tensor formed from bilinears of the
fundamental field  that can play a role identical to that  of the primary
scalar fields in previously investigated induced theories. A master field is
formed from the eigenvalues of this positive definite self-adjoint matrix. The
density function for the components of the master field satisfies a singular
integral equation that can, in particular cases, be solved exactly or in a
convergent series of  approximations.

 \section{Lattice
model for gauged fundamental scalars}

 Following the philosophy of induced gauge
theories the Wilson plaquette term is omitted from a lattice action in
the hope that correct gauge interactions are  generated though scalar
self-interactions. Then the only term of the Lagrangian in which the gauge
field makes its appearance is the gauged kinetic term.  The kinetic
terms written in terms of fundamental representation scalars,
$\psi^i_\alpha(x)$, and  compact gauge fields, $U_\mu(x)$, has the form
\begin{equation} L_K = \kappa\sum_{\mu=1}^D\left[\psi^\dagger_i(x)U_\mu(x)
\psi^i(x+a\hat e_\mu) + \psi^\dagger_i(x+a\hat e_\mu)U^\dagger_\mu(x)
\psi^i(x)\right], \label{kinetic} \end{equation} where $\kappa$ is the hopping
parameter, $a$ is the lattice constant. $i$ is a flavor index, while color
indices are suppressed. $D$ is the number of dimensions and $\hat e_\mu$ is a
unit vector along the $\mu$th coordinate axis.

  The hopping parameter can be eliminated by rescaling the scalar
field and absorbing the change into the coefficients of the scalar potential.
For that  reason in what follows we will set $\kappa=1$.

The Lagrangian also contains interaction terms. We only introduce $\psi^4$
interaction terms but other terms relevant or marginal in less than
four dimensions could also be introduced and treated without additional
difficulty. In principle there are two fourth order self interaction terms
allowed by the $U(N_c)\times U(N_f)$ symmetry, differing in their  symmetry
structure.  These terms are formed from pairs of bilinears of the fundamental
field  that are global scalars and gauge tensors or scalars, or alternatively
formed from gauge scalars that are global tensors or scalars. They have the
form \begin{equation}
L_I=L_T+L_S=\frac{g_T}{N_f}\psi^\dagger_i(x)\psi^j(x)\psi^\dagger_j(x)
\psi^i(x)+
\frac{g_S}{N_f^2}\psi^\dagger_i(x)\psi^i(x)\psi^\dagger_j(x)\psi^j(x).
\label{selfL} \end{equation} Terms with scalar coupling are not generated in
the
hopping parameter expansion. Thus, it is self-consistent to
keep the term with the tensor structure only. Therefore, in what follows we
will
 omit the scalar term and use the notation $g_T=g$.

The complete Lagrangian of the induced theory has then the form:
$
L=L_K+L_T+L_M,$ where $L_M$ is the mass term,
\begin{equation}
L_M = m^2\psi^\dagger_i(x)\psi^i(x).
\label{mass}
\end{equation}

Following Araf'eva~\cite{arefeva} we use an  Eguchi-Kawai type
realization~\cite{eguchi} of coordinate space in the internal symmetry space.
Such a reduction is admissible when the number of colors
and flavors is large.
In the large $N_f$ and
$N_c$ limit one can realize the spacetime translation operators on the
symmetry
groups as
$D_\mu(k)$ and $\Delta_\mu(p)$, where $D_\mu(k)$ is a diagonal $U(N_c)$
matrix,
with matrix elements $e^{iak^\alpha_\mu}$, $1\leq \alpha\leq N_c$, and
 $\Delta_\mu(p)$ is a diagonal $U(N_f)$ matrix,
with matrix elements $e^{iap^i_\mu}$, $1\leq i\leq N_f$. These operators
translate objects in the fundamental representation of the $SU(N_c)$
[$SU(N_f)$] group by one lattice unit along the $\mu$th axis. The model thus
obtained called "quenched" because the integration over the momentum variables
should be performed after physical amplitudes are calculated. Then
the lagrangian is reduced to a single site form
\begin{equation}  L =
m^2\psi^\dagger_i\psi^i + \frac{g}{N_f}\psi^\dagger_i\psi^j
\psi^\dagger_j\psi^i
+ \sum_{\mu}{\rm
Tr}\left[\Delta_\mu(p)\psi^\dagger_iD_\mu^{1/2}(k)U_\mu
D_\mu^{1/2}(k)\psi^i+ h.c\right], \label{lagrange2} \end{equation}
where, for the sake of symmetry, we assume that the gauge fields occupy sites
at the middle of links, halfway between two vertices of the lattice, where
the scalar particles reside. It can be shown that the perturbation expansion
of
this model coincides with that of the original model at large $N_f$ and $N_c$.

At this point it can be easily seen that the translation operators can be
eliminated from the theory altogether making the theory effectively
translation invariant.  The operator $\Delta_\mu(k)$ can be eliminated,
because
in gauge invariant closed loops its contributions cancel. The operators
$D_\mu(k)$ can be eliminated using a redefinition of the gauge field. Then we
arrive at the lagrangian
\begin{equation}
L = m^2\psi^\dagger_i\psi^i + \frac{g}{N_f}\psi^\dagger_i\psi^j
\psi^\dagger_j\psi^i +
\sum_{\mu}\left[\psi^\dagger_iU_\mu
\psi^i+ h.c\right], \label{lagrange3} \end{equation}
which reproduces the correct perturbation expansion.

The theory defined by Lagrangian (\ref{lagrange3})  fails in the weak coupling
limit because it does not reproduce the correct eigenvalue spectrum of the
matrices $U_\mu$. Therefore, one should introduce, along with Gross and
Kitazawa~\cite{gross1} a quenched theory, containing a gauge invariant
modification of the measure constraining the eigenvalues of $U_\mu$. The
appropriately modified theory will be discussed briefly at the end of this
paper and in more details in a future publication.  Throughout this paper we
will concentrate on the simple theory  defined by Lagrangian
(\ref{lagrange3}).

 At this point there is still no obvious candidate for a
master field. Notice, however that lagrangian (\ref{lagrange3}) depends only
on
the positive self adjoint tensor formed from bilinears of the scalar field, $
\Phi$ \begin{equation}
\Phi_\alpha^\beta=\psi^i_\alpha \psi^{\beta\dagger}_i,
\label{bilinear}
\end{equation}
where, for the sake of clarity the gauge indices $\alpha$ and $\beta$ are
displayed.

We now change variables by introducing the
following factor into the partition function
\begin{equation}
1=\int d\Phi \delta(\Phi-\psi^i\psi_i^\dagger)= \int d\Phi d\rho e^{{\rm Tr}
 \rho
(\Phi-\psi^i\psi_i^\dagger)}.
\label{substitution}
\end{equation}

Integrating  over the complex scalar fields we obtain
  \begin{equation}
\int d\psi^i = \int d\Phi d\rho e^{{\rm Tr}[ \rho
\Phi-N_f\log\rho]}.
\label{sub2}
\end{equation}
At large $N_f$ we can evaluate the the integral over $\rho$ using the saddle
point method.  Thus, to terms of
$O(1/N_f)$ we obtain
\begin{equation}
\int d\psi^i \sim \int d\Phi  e^{N_f{\rm Tr}\log \Phi}.
\label{sub3}
\end{equation}

We are now in the position of writing down the complete action in terms of the
fields  $\Phi$ and $U$. Action $S$ will be defined with a sign opposite to
the usual one so that $S$=log$Z$, where $Z$ is the Boltzmann factor. We
obtain \begin{equation}
S = {\rm Tr}\left[N_f\ \log \Phi- m^2\ \Phi - \frac{g}{N_f}\ \Phi^2 +
\sum_{\mu}\Phi\ (U_\mu
+ U_\mu^\dagger)\right]. \label{action1}
\end{equation}

The partition function is invariant under a unitary transformation of the
positive hermitian matrix, $\Phi$, $\Phi\rightarrow V\Phi V^\dagger$, that
diagonalizes $\Phi$.  The action, $S$, written in terms of the eigenvalues of
$\Phi$, $\phi_\alpha$, $\alpha=1,...,N_c$ is
\begin{equation}
S =
\sum_\alpha\left[\frac{1}{2}\sum_{\beta\not=\alpha}
\log(\phi_\alpha-\phi_\beta)^2+N_f
\log \phi_\alpha- m^2\phi_\alpha -
\frac{g}{N_f} \phi_\alpha^2\right] + D\ W(\phi),
\label{action7}
\end{equation}
where
\begin{equation}
W(\phi)= \log \int dU e^{  \Phi(U
+ U^\dagger)}
\label{def}
\end{equation}
is the single link integral studied  by Brower, Nauenberg,
Brezin and Gross~\cite{brower}\cite{brezin} in the large $N$ limit, and
$D$ is the number of dimensions. The first term of the action is the logarithm
of the Jacobian of the transformation from $\Phi$ to its eigenvalues,
$\phi_\alpha$.

As it will soon become clear the action is of $O(N_f^2)$. Then the most
probable eigenvalues can be calculated by using the saddle point method. The
values of $\phi_\alpha$ that maximize the action form the master field.  By
definition, all eigenvalues, $\phi_\alpha$, are positive. Zero  eigenvalues
are
excluded by the term $\log \Phi$ of the action.  At large $N$ these
eigenvalues
approximate a continuous distribution.
 As  usual, we introduce the density of eigenvalues, $\rho(\nu)$, by the
relation \begin{equation}
\frac{1}{N_f}\sum_{\alpha=1}^{N_c}\rightarrow \int d\nu \rho(\nu),
\label{sub}
\end{equation}
where $\nu$ is the running eigenvalue.  Consequently, $\rho(\nu)$ satisfies
the sum rule
\begin{equation}
\int \rho(\nu) d\nu= \epsilon=\frac{N_c}{N_f}.
\label{sumrule}
\end{equation}
The range of
$\nu$ is yet unknown. The positivity of matrix $\Phi$ requires that it starts
 at
some
$\nu_{\rm min}>0$.

After rescaling variable $\nu$  by the relation $\nu\rightarrow\nu N_f$, $N_f$
can be eliminated and the action will have a nontrivial dependence on
 $\epsilon$
only. Using the results of~\cite{brower}
\cite{brezin} and neglecting trivial constant terms we can write the action as
\begin{equation}
\frac{S}{N_f^2} =  \int d\nu \rho(\nu)\left[ \frac{1}{2}\int
d\lambda ~\rho(\lambda)\log(\nu-\lambda)^2+
\log\nu - m^2\nu
-g\nu^2\right]+D\ W,
\label{cont_ax}
\end{equation}
where
\begin{equation}
W = \int d\nu \rho(\nu)\left\{2(\nu^2+c)^{1/2}- \frac{1}{2}\int d\lambda
~\rho(\lambda)\log\left[(\nu^2+c)^{1/2}+(\lambda^2+c)^{1/2}\right]\right\}-c
\label{prop}
\end{equation}
The constant $c$ is determined from the equation
\begin{equation}
\int d\nu \frac{\rho(\nu)}{(\nu^2+c)^{1/2}}=2.
\label{c_det}
\end{equation}
if condition
\begin{equation}
s=\int d\nu \frac{\rho(\nu)}{\nu}>2.
\label{c_det2}
\end{equation}
is satisfied. This condition defines the strong coupling phase.  If $s<2$ then
$c=0$ and the system is in the weak coupling phase of the single site
integral.

 An integral equation for the optimal density distribution is
obtained if one varies (\ref{cont_ax}) with respect to the distribution
function
$\rho(\nu)$
\begin{equation}
\int d\mu \rho(\mu) \left[\log(\mu-\nu)^2-D
\log(\sqrt{\nu^2+c}+\sqrt{\mu^2+c})\right]+ \log\nu - m^2\nu+2
D\sqrt{\nu^2+c} -g\nu^2=\lambda ,
\label{new-s}
\end{equation}
where $\lambda$ is a variational parameter, corresponding to constraint
(\ref{sumrule}).

The derivative of (\ref{new-s}) with respect to $\nu$ provides a singular
integral equation that one hopes to solve for $\rho(x)$,
\begin{equation}
 P\int d\mu
\rho(\mu)\left(\frac{2}{\nu-\mu}-\frac{\nu}{\sqrt{\nu^2+c}}\frac{D}{\nu+\mu}
\right)
+\frac{1}{\nu}-m^2 +2 D \frac{\nu}{\sqrt{\nu^2+c}}-2 g \nu=0.
\label{int-eq}
\end{equation}

The solution of integral equation (\ref{int-eq}) does not determine the range
of eigenvalues (variable $\nu$). In fact, one can get a solution for all
choices of $0<\nu_{min}<\nu_{max}$.  We need to impose an additional
constraint,
$\rho(\nu)>0$,  to get a unique solution. The positivity of
$\rho(\nu)$, as the density of eigenvalues is fairly obvious. Formally,
it is required by the uniqueness of eigenvalues $\phi_\alpha$ of matrix
$\Phi$ as
a function of $\alpha$.

In principle we could obtain equations for  $\nu_{min}$ and $\nu_{max}$ by
varying the action with respect to these parameters.  Being dependent on
$d\nu \rho(\nu)$ only, these variations  lead to  (\ref{new-s}), taken at the
upper or lower limits and multiplied by the value of $\rho(\nu)$ taken at the
same points.  It follows then that the variation of the action has a double
zero
at a  density
$\rho(x)$ satisfying the variational equation and also vanishing at the end
points. Consequently, the action has a point of inflection as a function
of the
end point parameters. If one slightly varies $\nu_{min}$ (or $\nu_{max}$)
from
the points at which $\rho(\nu_{min})=\rho(\nu_{max})=0$, with intention to
{\em increase $S$}, then the density function becomes negative near the
end point, leading to an inadmissible solution. Consequently, one must
demand
the vanishing of the density function $\rho(\nu)$ at the end points.
As we will
see later this condition determines the end points uniquely.

The method we use for determining the range of variation of eigenvalues
can be
applied to a wide range of problems involving matrix variables in the
large $N$
limit and master fields.  If the action is of the form
\begin{equation}
S=\sum_k \prod_{i=1}^k\left[\int_{\nu_{min}}^{\nu_{max}} {\rm d}\nu_i\
\rho(\nu_i)\right] F_k(\nu_1,...,\nu_k),
\label{generic}
\end{equation}
then the spectral function must vanish at the boundaries, $\nu_{min}$
and $\nu_{max}$.  In turn, this requirement can be used to determine
$\nu_{min}$
and $\nu_{max}$.

We have not been able
to find an exact solution of (\ref{int-eq}) at arbitrary $D$,
either in the weak or in the strong coupling region.  At $D=2$ an exact
solution can be found in the weak coupling phase ($c=0$). Furthermore, at
$D>2$
the integral equation can be transformed into a Fredholm type equation
providing
a {\em convergent}  expansion for $\rho(\nu)$, displaying critical behavior,
around $D=2$ and a, probably continuous, phase transition as a function of
$\epsilon$.  The solution of the Fredholm equation at
$D>2$ is also possible using a convergent expansion in a power series of
$\epsilon$.

\section{Exact solution of the model at $D=2$}

If the number of dimensions, $D=2$, then the problem of finding the density
function simplifies considerably  in the weak coupling phase of $W(a)$. To see
this let us consider (\ref{int-eq}) in the weak coupling phase
\begin{equation}
P\int d\mu \rho(\mu) \left[\frac{2}{\nu-\mu}-\frac{D}{
\nu +\mu}\right]+ \frac{1}{\nu} - M^2 -2\nu=0 ,
\label{newest-s}
\end{equation}
 where $M$ is the renormalized mass defined by $M^2 =
\frac{m^2 -2D}{\sqrt{g}}$.
By redefining the mass parameter we scale out the coupling constant,
$g$, from
action.  In the weak coupling phase it only appears in the rescaled relation
(\ref{c_det}) \begin{equation}
s=\sqrt{g}\int d\nu \frac{\rho(\nu)}{\nu}<2.
\label{new_c_det}
\end{equation}
By an appropriate choice of $g$ it is always possible to tune $s$ to an
arbitrary value inside the weak coupling phase, e.g to the boundary of the
strong and weak coupling phases, where $s=2$. Note that in the strong
coupling regime $c$ approaches zero continuously when the critical point
$s=2$ is approached. Therefore, if $s$ is tuned to the critical point
then our solution is valid, no matter which side the critical point is
approached from.

At $D=2$ the derivative of variational equation (\ref{newest-s}) can be
transformed to
\begin{equation}
P\int_{-1}^1 dx \frac{R(x)}{y-x} =-\frac{1}{\sqrt{A(1+ z x)}}+M^2 + 2
\sqrt{A(1+ z x)},
\label{d2vari}
\end{equation}
where the constants $A$, $z$ and variable $x$ have been introduced by the
relations
\begin{equation}
\nu=\sqrt{A\ (1+ z\ x)},  ~~~~~\nu_{max}=\sqrt{A\ (1+z)},
~~~~~\nu_{min}=\sqrt{A\ (1-z)}, .
\label{defs}
\end{equation}
and where we define $R(x)=\rho(\nu)$
In other words, the range of variable $x$ is $-1<x<1$.

Determining end
points $\nu_{max}$ and $\nu_{min}$ is equivalent to determining $A$ and $z$.
As
we will see later these parameters are fixed by the requirement of
$R(1)=R(-1)=0$.

A solution of integral equation (\ref{d2vari}) that vanishes at the end
points, $x=\pm1$, can be obtained by  inverting the Cauchy principal value
integral.  After obvious manipulations one obtains the following expression
for the density function  \begin{eqnarray}
R(x)& =&-\frac{\sqrt{1-x^2}}{4\ \pi^2} P\int_{-1}^1 \frac{{\rm
d}y}{(x-y)\ \sqrt{1-y^2}}\left(-\frac{1}{\sqrt{A\ (1+ z\ y)}}+M^2 + 2
\sqrt{A\ (1+ z\ y)}\right) \nonumber\\& = & \frac{\sqrt{A}\ z\
\sqrt{1-x^2}}{2\
\pi^2}
\int_{-1}^1\frac{2 {\rm d}\xi}{(\sqrt{1+\xi\ z}+\sqrt{1+x\
z})\sqrt{1-\xi^2}}\nonumber \\ &+&\frac{z\
\sqrt{1-x^2}}{4\
\pi^2\ \sqrt{A}\ \sqrt{1+ x\ z}}\int_{-1}^1 \frac{{\rm d} \xi}{(\sqrt{1+\xi\
z}+\sqrt{1+x\ z})\sqrt{1+\xi\ z}\sqrt{1-\xi^2}}.
\label{rexpr}
\end{eqnarray}
The density, $R(x)$, is obviously positive definite and as such, physically
admissible. We could  express $R(x)$ in a closed form as a combination of
complete elliptic functions. As it is not especially  illuminating that form
will not be presented in this paper.

 At this point $A$ and $z$ are yet undetermined.
 It is comparatively easy to see that consistency conditions
 allow us to determine
these constants. Before discussing these conditions it is worth mentioning
that solutions of integral equation (\ref{d2vari}), not vanishing at
$x=\pm1$, could be found at every choice of the end points. These solutions
would either not maximize $S$ or would not be positive definite.

To determine the end points $\nu_{min}$ and $\nu_{max}$ (i.e. parameters $A$
and $z$) first note that form (\ref{rexpr}) seems to be independent of
$M^2$. This is so because when the Cauchy integral is inverted the term $M^2$
on
the right hand side of  (\ref{rexpr}) gives a vanishing contribution. This
term
can however be recovered if we substitute expression (\ref{rexpr}) into
(\ref{d2vari}) and use the Poincare-Bertrand formula for the exchange of the
order of Cauchy integrals. Then all terms of the right hand side of
(\ref{d2vari}) except $M^2$ are recovered. A constant term appears that must
be
identified with $M^2$: \begin{equation} \frac{1}{\pi}\int_{-1}^1 \frac{{\rm
d}\xi}{ \sqrt{1-\xi^2}}\left(\frac{1}{\sqrt{A\ (1+ z\ \xi)}}- 2 \sqrt{A\
(1+ z\
\xi)}\right)= M^2. \label{rel1}
\end{equation}
This equation can be used to determine $A$ in terms of $z$ and $M$.

The second condition is obtained from the normalization of the spectral
function, (\ref{sumrule}).  After simple manipulations we obtain the
following relation:
\begin{equation}
-\frac{1}{2\ \pi}\int_{-1}^1 \frac{\xi\ {\rm d}\xi}{
\sqrt{1-\xi^2}}\left(\frac{1}{\sqrt{A\ (1+ z\ \xi)}}- 2
\sqrt{A\ (1+ z\ \xi)}\right)= \epsilon.
\label{rel2}
\end{equation}
Having determined $A$ as a function of $z$ and $M$ from (\ref{rel1}) we can
use  (\ref{rel2}) to determine $z$ as a function of the physical parameters
$M$ and $\epsilon$.

Numerical calculations show that one can find unique positive solutions for
$A$
and $z$ at all possible choices of real $M^2$ and positive $\epsilon$. As a
contrast, we will see later at $D>2$ we can find a similar type of solution
only in a restricted range of $\epsilon$.  $D=2$
is a lower critical dimension of the theory.

We now examine the behavior of the solution in the limits of small and large
$\epsilon$. At small values of $\epsilon=N_c/N_f$~
$z\sim\epsilon$. In other words,  the range of eigenvalues relative to the
average value shrinks to zero.  Then the matrix $\Phi$ tends to a constant
multiple of the unit matrix. We can also see by inspection that in the limit
$z\rightarrow0$ the distribution of eigenvalues follows Wigner's semicircle
law.

At large
$\epsilon$~~ $z\sim 1 -e^{-\epsilon^2}$, the lower bound of the eigenvalues of
matrix $\Phi$, $\nu_{min}=A\ (1-z)\sim \epsilon e^{-\epsilon^2}$, approaches
zero,
 while the
upper limit, $\nu_{max}=A\ (1+z) \sim 2\epsilon$, tends to infinity, so that
the
eigenvalues span the whole positive real axis. The distribution of
eigenvalues,
$R(x)$, differs substantially from the semicircle law, favoring eigenvalues
close to the lower bound.

 \section{Induced theories at $D>2$ }

At $D>2$ we have not been able to find an exact solution for the density
function.   It turns out however that even at $D>2$ the singular
integral equation can be transformed into a Fredholm equation, solutions of
which can be generated in convergent expansions.  Having these expansions is
almost as useful as having an exact solution.

 We will start again from the derivative of (\ref{newest-s}) with respect to
$\nu$. Using variable $x$ and parameters
$A$ and $z$ as at $D=2$, in (\ref{d2vari}) we obtain
 \begin{eqnarray} P\int_{-1}^1 dx
\frac{R(x)}{y-x}& =&\frac{(D-2)\ z}{2}\int_{-1}^1\frac{R(x)\ {\rm
d}x}{\sqrt{1+z\ x}\ (\sqrt{1+z\ x}+\sqrt{1 + z\
y})}\nonumber \\ &-&\frac{1}{\sqrt{A(1+ z x)}}+M^2 + 2 \sqrt{A(1+ z x)},
\label{dgreq2vari} \end{eqnarray} This equation is identical to
(\ref{d2vari})
other then the appearance of the first term on the right hand side.

 After inverting the Cauchy integral on the
left hand side of (\ref{dgreq2vari}) we obtain
\begin{eqnarray}  R(x)& =& f(x) +(D-2)\ \int_{-1}^1
{\rm d}y \ K(x,y)\ R(y)\nonumber \\ &\equiv &\frac{\sqrt{A}\ z\
\sqrt{1-x^2}}{2\
\pi^2} \int_{-1}^1\frac{2 {\rm d}\xi}{(\sqrt{1+\xi\ z}+\sqrt{1+x\
z})\sqrt{1-\xi^2}}\nonumber \\ &+&\frac{z\ \sqrt{1-x^2}}{4\
\pi^2\ \sqrt{A}\ \sqrt{1+ x\ z}}\int_{-1}^1 \frac{{\rm d} \xi}{(\sqrt{1+\xi\
z}+\sqrt{1+x\ z})\sqrt{1+\xi\ z}\sqrt{1-\xi^2}}\nonumber\\ &-&\frac{(D-2)\
z^2\ \sqrt{1-x^2}}{4\ \pi^2}\int_{-1}^1 \frac{R(y) \ {\rm d}y}{(\sqrt{1+y\
z}+\sqrt{1+x\ z})\ \sqrt{1+y\
z}}\nonumber\\ &\times&\int_{-1}^1\frac{{\rm d}\xi}{(\sqrt{1+\xi\
z}+\sqrt{1+x\ z})(\sqrt{1+\xi\
z}+\sqrt{1+y\ z})\sqrt{1-\xi^2}}.
\label{rexprdgreq2}
\end{eqnarray}
This is a linear, inhomogeneous integral equation with an
inhomogeneity that is identical to the density function at $D=2$,  with
kernel \begin{eqnarray} K(x,y)&=&-\frac{
z^2\ \sqrt{1-x^2}}{4\ \pi^2\ (\sqrt{1+y\
z}+\sqrt{1+x\ z})\ \sqrt{1+y\
z}}\nonumber\\ &\times& \int_{-1}^1\frac{{\rm d}\xi}{(\sqrt{1+\xi\
z}+\sqrt{1+x\ z})(\sqrt{1+\xi\
z}+\sqrt{1+y\ z})\sqrt{1-\xi^2}},
\label{kernel}
\end{eqnarray}
and with expansion parameter $D-2$.
Kernel $K(x,y)$ can easily be symmetrized. It is bounded, i.e. it satisfies
condition
 \begin{equation}
\int_{-1}^1 {\rm d}x\ \int_{-1}^1 {\rm d}y |K(x,y)|^2 < \infty.
\label{finite}
\end{equation}
It follows that (\ref{rexprdgreq2}) is a Fredholm equation.

The solution of the integral equation is given
by the following expression \begin{equation}
R(x)= f(x) + \int_{-1}^1 {\rm d}y \ Res(x,y)\ f(y),
\label{soln-fred}
\end{equation}
where $f(x)$ is the inhomogeneity of the integral equation,
and $Res(x,y)$ is the resolvent of the equation,
\begin{equation}
Res(x,y)=\frac{(D-2)\ K(x,y)}{1-(D-2)\ K(x,y)}=\frac{N(x,y)}{Det}.
\label{resolvent}
\end{equation}
The Fredholm numerator, $N(x,y)$, and the Fredholm determinant, $Det$,
are entire functions of the expansion parameter, $D-2$.  The first terms of
their expansions are given by
\begin{eqnarray}
N(x,y)&=& (D-2)K(x,y)-\frac{(D-2)^2}{2!}\left(K(x,y)\ {\rm
Tr}K-K^2(x,y)\right)+... \label{num}
\\ Det &=& 1 - (D-2) {\rm Tr} K +\frac{(D-2)^2}{2!}\left[ ({\rm
Tr}K)^2 -{\rm Tr} K^2\right],
\label{det}
\end{eqnarray}
where $K^2$ is understood in the operator sense.

The negative definite kernel insures that at $D>2$ the Fredholm determinant
has
no zeros. At zeros of the Fredholm determinant only the homogeneous equation
has a solution.  Thus,
the only reason why solutions of  (\ref{rexprdgreq2}) would not be acceptable
is
the lack of positivity at all $x$.

Note that at
$z=0$  the kernel vanishes. Then, in view of the positivity of the
inhomogeneity at all $-1<x<1$ the spectral function is physically acceptable.
Since small $z$ and small $\epsilon$ are synonymous, we find that at
all $D>2$
one can find a critical value of $\epsilon$, such that for
$\epsilon< \epsilon_c(D)$ a physically acceptable solution exists. As we saw
earlier  $\epsilon_c(2)=\infty$.

 By examining the solution at $D=2$ one can see that  $z\rightarrow1$
corresponds to $\epsilon\rightarrow\infty$. Therefore, one would expect
that at
$D>2$ limitations on the range  of admissible values of $z$ may appear near
$z=1$
(large $\epsilon$), excluding large values of $\epsilon$. Therefore we
performed a systematic expansion of the Fredholm determinant and solution
$R(x)$
near $z=1$.

Due to singularities at $x=-1$ the terms of $Det$ (and of $N(x,y)$) diverge
logarithmically near $z=1$.  The Fredholm determinant can be written as
\begin{equation}
Det= \exp\left\{{\rm Tr} \log (1-(D-2)\ K)\right\}=\exp\left\{-(D-2)\ {\rm
Tr}K-\frac{(D-2)^2}{2} {\rm Tr} K^2 +...\right\}.
\label{det1}
\end{equation}
It is easy
to calculate the leading (and even subleading) logarithms of the traces of
powers of the kernel. Using these we obtain the following expression for the
Fredholm determinant near $z=1$:
\begin{equation}
Det= (1-z)^{\alpha(D)},
\label{singdet}
\end{equation}
where
\begin{equation}
\alpha(D)=
\frac{1}{\pi^2}\left[\frac{D-2}{2}-\frac{(D-2)^2}{3}+O((D-3)^3)\right].
\label{expandalpha} \end{equation}
This is a critical exponent.   Due to the rearrangement of the Fredholm
series, this expansion, as similar perturbation expansions, is
asymptotic only.

The relation of parameters $A$ and $z$ to $M^2$ and $\epsilon$ is similar to
relations (\ref{rel1}) and (\ref{rel2}):
 \begin{eqnarray}
&&\frac{1}{\pi}\int_{-1}^1 \frac{{\rm d}\xi}{
\sqrt{1-\xi^2}}\Big(\frac{1}{\sqrt{A\ (1+ z\ \xi)}}- 2
\sqrt{A\ (1+ z\ \xi)}\nonumber \\ &-&\frac{(D-2)\ z}{2}\int_{-1}^1
\frac{R(y)\
{\rm d}y}{\sqrt{ 1+ z\ y}\ (\sqrt{ 1+ z\ \xi}+\sqrt{ 1+ z\ y})}\Big)= M^2.
\label{rel21} \end{eqnarray}
and
\begin{eqnarray}
&-&\frac{1}{2\ \pi}\int_{-1}^1 \frac{\xi\ {\rm d}\xi}{
\sqrt{1-\xi^2}}\Big(\frac{1}{\sqrt{A\ (1+ z\ \xi)}}- 2
\sqrt{A\ (1+ z\ \xi)}\nonumber\\ &-&\frac{(D-2)\ z}{2}\int_{-1}^1 \frac{R(y)\
{\rm d}y}{\sqrt{ 1+ z\ y}\ (\sqrt{ 1+ z\ \xi}+\sqrt{ 1+ z\ y})}\Big)=
\epsilon.
\label{rel22}
\end{eqnarray}

We will investigate now the  positivity of the density
function. It is fairly easy to show that in the immediate neighborhood of
$x=-1$, where the kernel and inhomogeneity become singular at $z=1$, the
iterated density function is positive definite.  Therefore we will
concentrate on the region of $x+1>>1-z$.

The first nontrivial approximation to
the Fredholm series is identical to the once iterated solution of the integral
equation.  It provides an asymptotic relation for the range of $\epsilon$. To
see this we investigate the dependence of $A$ on $z$.  This must be done
before
one is able to calculate the leading contributions to the density function.
If we study solutions at vanishing or small renormalized mass. $M$, then
the location of the critical point is independent of $M$ in leading orders of
the expansion.

We calculated nonleading order
corrections, as well, just to see how much they influence asymptotic
relationships. Using (\ref{rel21}) we obtain \begin{equation}
A=\log\left(\frac{2}{1-z}\right) \frac{1-(D-2)\log[2/(1-z)]/192
\pi^2}{8+0.82247 \ (D-2)/\pi^2} \label{aequ} \end{equation}
 When the leading asymptotic part of this relation is substituted into the
expression of the density function
  we obtain
\begin{equation} R(x)\simeq \frac{\sqrt{1-x^2}}{(1 +
z\ x)\ 2\ \sqrt{2}\ \pi^2\ \sqrt{A}}\ \log\left(\frac{2}{1-z}\right) \
\left(1-\frac{(D-2)\ \log^2 \left(\frac{2}{1-z}\right)}{12\ \sqrt{2}\ \pi^2}
\right)  .
\label{iterated}\end{equation}

The density function becomes indefinite
 at large enough  $D-2$. This is a nonsingular structural phase transition,
except in the vicinity of $D=2$.  Near the transition $(D-2)\ \left(
\log[2/(1-z)]\right)^2=O(1)$, so at large $\log[2/(1-z)]$ the corrections in
(\ref{aequ}) are indeed small.

Having determined the critical value of $z$ (the maximal $z$ with admissible
spectral function) we can determine the critical value of $\epsilon$ using
(\ref{rel22}). Substituting into this relation we get
\begin{equation}
\epsilon= \sqrt{\log\left[\frac{2}{1-z}\right]} \ \frac{4}{3\ \pi}+...
\label{eps}
\end{equation}
Combining (\ref{eps}) and (\ref{iterated}) we get the final expression for the
critical value of $\epsilon$
\begin{equation}
\epsilon_c=\frac{1}{(D-2)^{1/4}}\ \left(\frac{1024\ \sqrt{2}}{27\
\pi^2}\right)^{1/4}= \frac{2.331}{(D-2)^{1/4}}.
\label{epscrit}
\end{equation}
As expected, at $D=2$ the critical value of $\epsilon $ diverges. The
prediction of (\ref{epscrit}) for the value   of $\epsilon_c(D)$  is not
reliable at $D=4$.  A better value could be obtained from a thorough numerical
study.

It is interesting to speculate what happens at $\epsilon>\epsilon_c(D)$. It is
easy to see that density functions, satisfying $\rho(\nu)>0$ and having finite
action still exist. The positive density function that maximizes $S$ at
$\epsilon>\epsilon_c$ may not belong to the class of density functions we have
been investigating above.  One is able to get a hint about the nature of the
density function in this region if one considers that when
$\epsilon\rightarrow\epsilon_c(D)-0$  then $R(x)$ develops a zero at some
intermediate value of $x$, $x=x_c$, where $-1<x_c<1$.   Then  at
$\epsilon>\epsilon_c(D)$ the spectrum of eigenvalues should split into two
disjunct intervals, the density function vanishing at the end points of both
intervals. This would be a density function that does not belong to the class
of functions considered above. Clearly, the change from one to the other type
of density functions is, an almost probably continuous, structural phase
transition.

 To study the nature of these phases it would be very important
to derive the dependence of large Wilson loops, \begin{equation}
W[X,Y]= \langle {\rm Tr}\left[(U_x)^X\ (U_y)^Y\ (U_x)^{-X}\ (U_y)^{-Y}\right]
\rangle.
\label{wilson}
\end{equation}
on the area of the loop, $XY$.
This, and the investigation of the model in the strong coupling regime of the
single link integral are left to a future publication.

\section{Quenched Theory}

In the above described theory, in the weak coupling phase,  the
distribution of
eigenvalues of the gauge matrices is incorrect.  To
eliminate this problem one must constrain the eigenvalues of the gauge
field $U_\mu$ in a gauge invariant manner. One possible approach is that of
Gross and Kitazawa~\cite{gross1} that introduces the following
modification of
the measure:
\begin{equation} \int dU_\mu \rightarrow \int dU_\mu \prod_\mu
\int dV_\mu \Delta(D_\mu)  \delta(D_\mu^{1/2}U_\mu D_\mu^{1/2} - V_\mu D_\mu
V_\mu^\dagger). \label{substitute}
\end{equation}
While this addition to the measure  fixes the eigenvalue distribution of
matrices $U_\mu$, the substitution does not affect agreement with the
perturbation expansion of the original theory.

Substituting into the Lagrangian we obtain
\begin{equation}
L = m^2\psi^\dagger_i\psi^i + \frac{g}{N_f}\psi^\dagger_i\psi^j
\psi^\dagger_j\psi^i +
\kappa\sum_{\mu}\left[\psi^\dagger_iV_\mu
(D_\mu + D_\mu^\dagger) V_\mu^\dagger\psi^i\right], \label{lagrange4}
\end{equation}

At this point one can introduce the bilinear combinations just as in
(\ref{bilinear}) to get the quenched theory for the positive adjoint matrix
$\Phi$, The action, in terms of the eigenvalues of $\Phi$, $\phi_\alpha$
takes
the form \begin{equation}
S =
\sum_\alpha\left[\frac{1}{2}\sum_{\beta\not=\alpha}
\log(\phi_\alpha-\phi_\beta)^2+N_f
\log \phi_\alpha- m^2\phi_\alpha - \frac{g}{N_f} \phi_\alpha^2\right] +
\sum_{\mu}\log I(\Phi, C_\mu), \label{lagrange7}
\end{equation}
where $I(\Phi,C_\mu)$ is the Itzykson-Zuber integral~\cite{zuber}
\begin{equation}
I(\Phi, C_\mu) = \frac{{\rm Det}\left[e^{2\kappa \phi_\alpha \cos
k_\mu^\beta}\right]}{\Delta(\phi_\alpha)
\Delta(\cos k_\mu^\beta)}= \int dV_\mu e^{2\kappa\sum_{\mu}{\rm Tr}\Phi
V_\mu
C_\mu V_\mu^\dagger},
\label{def2}
\end{equation}
and where
\begin{equation}
\Delta(a_\alpha) = \prod_{\alpha\not=\beta}(a_\alpha-a_\beta)
\label{delta}
\end{equation}
and $C_\mu$ is a diagonal matrix formed from the cosines of the momentum
variables.

The  density function for the distribution of continuous eigenvalues can
proceed the same way as it was done in the previous sections. The singular
integral equation derived  from the variation of the action is complicated
but
can be studied in the weak and strong coupling limits of the Itzykson-Zuber
integral.

Finally, we present the expression of Wilson loops in terms of the variables
used in this section:
 The Wilson loop of edges
$X$ and
$Y$ along the
$x$ and
$y$ axes is defined in the quenched theory as
\begin{eqnarray}
\langle W(X,Y)\rangle&=& {\rm Tr}\left[(D_x^{1/2}U_x
D_x^{1/2})^X(D_y^{1/2}U_\nu D_y^{1/2})^Y(D_x^{1/2}U_x
D_x^{1/2})^{-X}(D_y^{1/2}U_\nu D_y^{1/2})^{-Y}\right]\nonumber \\
&=&{\rm Tr}\left[V_x e^{iXk_x}V^\dagger_xV_ye^{iYk_y}V^\dagger_y V_x
e^{-iXk_x}V^\dagger_xV_ye^{-iYk_y}V^\dagger_y\right]
\label{wilson-q}
\end{eqnarray}
\section{Conclusion}
The investigation of gauge theories induced by scalars in the fundamental
representation may solve some of the problems that have arisen in  theories
induced by adjoint representation scalars.  Prominent among these is the
vanishing of the expectation value of the Wilson loop.  This is the main
reasons why we started an investigation of these theories.

The most serious difficulty in solving theories induced by fundamental
representation scalars has been the apparent lack of candidates for a master
field.  We solved this problems by showing  that bilinears formed
from fundamental representation fields form self adjoint matrices the
eigenvalues of which are the perfect candidates for a master field.

Using an Eguchi-Kawai type reduced representation we obtained an integral
equation for the density of eigenvalues that could be solved  exactly in two
dimensions at all values of the physical parameters. At $D>2$ we obtained a
Fredholm integral equation for the density function.  Using the perturbative
solution of this equation we established that the spectrum undergoes a phase
transition at a critical value of $\epsilon=N_f/N_c$.
Further investigations are needed to examine the nature of the phase
transition
at $D>2$ and the size dependence of the expectation value of Wilson loops.

The translation invariant approximation used in most of this paper
should be complemented by a non-translation invariant approach.
Investigation of theories induced by fundamental scalars using the
Gross-Kitazawa quenching prescription~\cite{gross1} is the next step of
investigations.

\begin{center} {\bf Acknowledgment} \end{center} This work was
supported in part by the U.S. Department of Energy Grant \#DE-FG02-84ER40153.

 \end{document}